\documentclass[12pt]{article}

\usepackage{amsfonts,amssymb,amsmath,amsthm,epsfig}
\usepackage{diagrams,cite}

\newcommand\nts{\negthickspace}
\newcommand\bns{\nts \nts \nts}
\newcommand\MM{\mathcal{M}}
\newcommand\dM{\partial \MM}

\newcommand{\beqs}{\begin{equation*}}
\newcommand{\beq}{\begin{equation}}

\newcommand{\eeqs}{\end{equation*}}
\newcommand{\eeq}{\end{equation}}

\newcommand{\beqas}{\begin{eqnarray*}}
\newcommand{\beqa}{\begin{eqnarray}}

\newcommand{\eeqas}{\end{eqnarray*}}
\newcommand{\eeqa}{\end{eqnarray}}

\newcommand{\eq}[2]{\begin{equation} #1 \label{#2} \end{equation}}

\newcommand{\eps}{\varepsilon}
\newcommand{\al}{\alpha}
\newcommand{\be}{\beta}
\newcommand{\ga}{\gamma}
\newcommand{\de}{\delta}

\newcommand{\ka}{\kappa}
\newcommand{\la}{\lambda}
\newcommand{\si}{\sigma}

\newcommand{\Ga}{\Gamma}

\newcommand{\Om}{\Omega}

\newcommand{\La}{\Lambda}

\newcommand{\blist}{\begin{itemize}}

\newcommand{\elist}{\end{itemize}}

\providecommand{\href}[2]{#2}

\DeclareFontFamily{OT1}{rsfs}{}
\DeclareFontShape{OT1}{rsfs}{m}{n}{ <-7> rsfs5 <7-10> rsfs7 <10->rsfs10}{} 
\DeclareMathAlphabet{\mycal}{OT1}{rsfs}{m}{n}

\DeclareMathOperator{\extdm}{d}
\newcommand{\extd}{\extdm \!}

\begin{document}

\begin{titlepage}
{\hfill MIT/CTP-3925}
\begin{center}

\vspace{7ex}

\textbf{\Large 
%WORKING NOTES\\ \vspace{4ex}
Liouville gravity from Einstein gravity\footnote{Platinum Jubilee, Indian Statistical Institue, Kolkata, India}
}\\

\vspace{7ex}

D.~Grumiller and R.~Jackiw

\vspace{4ex}

{\em Center for Theoretical Physics, Massachusetts Institute of Technology, \\
77 Massachusetts Ave., Cambridge, MA 02139, USA}

\vspace{4ex}

\end{center}
\vspace{7ex}
\begin{abstract}

%We consider Einstein gravity in $2+\eps$ dimensions, spherically reduce to $2$ dimensions, dualize, take the limit of vanishing $\eps$, dualize back and obtain Liouville gravity. We perform several consistency checks of our procedure: geometric issues, interactions with matter and Bekenstein-Hawking entropy. 

We show that Liouville gravity arises as the limit of pure Einstein gravity in $2+\eps$ dimensions as $\eps$ goes to zero, provided Newton's constant scales with $\eps$. Our procedure -- spherical reduction, dualization, limit, dualizing back -- passes several consistency tests: geometric properties, interactions with matter and the Bekenstein-Hawking entropy are as expected from Einstein gravity.

\end{abstract}
\end{titlepage}

\section{Introduction}\label{se:1}

Gravity in or near two dimensions has many manifestations. 
Gravity in $2+\eps$ dimensions serves as a toy model for quantum gravity and is known to be asymptotically safe \cite{Weinberg:1979}. 
The limit of the Einstein-Hilbert action,
\eq{
\lim_{\eps\to 0} I_{2+\eps}=\lim_{\eps\to 0}-\frac{1}{\kappa_{2+\eps}} \int \extd^{2+\eps} x \sqrt{|g|} \,R \,,
}{eq:2de1}
essentially yields the Euler characteristic, scaled by the gravitational coupling constant $\kappa_{2+\eps}=16\pi G_{2+\eps}$, where $G_{2+\eps}$ is Newton's constant. Taking into account 1-loop effects makes the limit $\eps\to 0$ non-trivial, because the relation between bare and renormalized coupling involves a term $1/\eps$ \cite{Weinberg:1979,Gastmans:1977ad,Christensen:1978sc,Kawai:1992fz}.

Also gravity in two dimensions serves as a toy model for quantum gravity and black hole evaporation. The first formulation of 2-dimensional gravity is due to Jackiw and Teitelboim (Bunster) \cite{Jackiw:1984,Teitelboim:1984}. Since then many similar models have been introduced, for instance the CGHS model \cite{Callan:1992rs} or the Witten black hole \cite{Witten:1991yr}. All of them are special cases of general dilaton gravity 
\eq{
I_{\rm 2dg} = - \frac{1}{\hat{\ka}}\,\int \extd^2 x \sqrt{|g|} \, \left(\phi R-U(\phi)\,(\nabla\phi)^2-V(\phi)\right) \,.
}{eq:2de17}
Here $\phi$ is a scalar field, the dilaton. For a review cf.~e.g.~\cite{Grumiller:2002nm}. A comprehensive list of potentials $U$ and $V$ can be found in \cite{Grumiller:2006rc}.

A related 2-dimensional gravity model is Liouville gravity, 
\eq{
I_{\rm L}^{(b)} = \frac{1}{4\pi} \int \extd^2x \sqrt{|g|}\,\big(Q\,\Phi R +(\nabla\Phi)^2+4\pi\mu e^{2b\Phi}\big)\,,
}{eq:2de33}
where $Q=b+b^{-1}$ and $b$, $\mu$ are constant. 
The fields $\Phi$ and $g$, as well as the action \eqref{eq:2de33}, may have various interpretations. If metric and scalar field are dynamical fields then the Liouville model \eqref{eq:2de33} is a special case of dilaton gravity \eqref{eq:2de17} with constant $U$ and exponential $V$. This is the case of relevance for our present work.\footnote{We shall recall another interpretation arising in the context of conformal field theory and string theory \cite{Ginsparg:1993is,Nakayama:2004vk} in the body of this paper.}
An interesting limit of \eqref{eq:2de33} is obtained as $b$ tends to zero. By virtue of the redefinition $\Phi=-Q\phi$ the action \eqref{eq:2de33} in the limit of $b\to 0$ can be brought into the form
\eq{
I_{\rm L} = -\frac{1}{\hat{\ka}} \int \extd^2x \sqrt{|g|}\,\big(\phi R - (\nabla\phi)^2 - \lambda e^{-2\phi}\big)\,,
}{eq:2de34}
where $\lambda = 4\pi\mu/Q^2$ and $\hat{\ka}=4\pi/Q^2$ have to be rescaled in such a way that they remain finite in the limit. The action \eqref{eq:2de34} is recognized as a special case of \eqref{eq:2de17} ($U_L=1$, $V_L=\la e^{-2\phi}$) and coincides with the Liouville theory studied e.g.~in \cite{Jackiw:2005su} for its Weyl transformation properties. 

It is the purpose of this paper to consider the limit in \eqref{eq:2de1}, but rather than keeping Newton's constant fixed we scale it such that $\kappa_{2+\eps}\propto\eps$. Because the limiting action \eqref{eq:2de1} is effectively vanishing as far as bulk properties are concerned, this rescaling of $\kappa_{2+\eps}$ leads to an indeterminancy of the form $0/0$, which is capable to yield an interesting bulk action. However, to make sense of such a limit we need something like a l'Hospital rule.
In order to establish such a rule we are guided by the following observation: Einstein gravity in $D$ dimensions exhibits $D(D-3)/2$ graviton modes, yielding at $D=2$ a negative number of graviton modes, which is difficult to interpret. It would be more convenient if, as $D$ is varied to $2$, the traversed number of degrees of freedom were positive and did not change continuously. We can achieve this by restricting to the s-wave sector of Einstein gravity, because it exhibits precisely zero propagating physical degrees of freedom, regardless of the dimension. Spherical symmetry implies the existence of $(D-2)(D-1)/2$ Killing vectors, which certainly is a strong restriction for large $D$. However, as $D=2$ is approached the number of Killing vectors required for spherical symmetry drops to zero, so our Ansatz of restricting to the s-wave sector does not lead to any symmetry constraints on the limiting geometry. This makes it plausible that our procedure captures all essential features.

The convenient trick of restricting to the s-wave sector is not sufficient to establish a meaningful limit $\eps\to 0$, but it allows to exploit properties unique to 2-dimensional dilaton gravity \eqref{eq:2de17}. In particular, we shall employ a certain duality \cite{Grumiller:2006xz} that renders the limit well-defined. Because the duality is involutive we shall dualize back after taking the limit and obtain in this way a non-trivial limit of \eqref{eq:2de1}, which turns out to be the Liouville action \eqref{eq:2de34}.

Our work is organized as follows. In Section \ref{se:2} we establish the l'Hospital rule as outlined above and obtain the limiting action. We perform several consistency checks in Section \ref{se:4}: we demonstrate that the geometric properties are reasonable, that our limiting action is consistent with the standard folklore that ``matter tells geometry how to curve'', and that the Bekenstein-Hawking entropy is consistent with the scaling behavior of Newton's constant. We  consider boundary terms and summarize in Section \ref{se:5}.

\section{Constructing the limiting action}\label{se:2}

We start with the Einstein-Hilbert action in $D$ dimensions\footnote{We omit boundary terms for the time being, because they can be constructed unambiguously once the limit of the bulk action is known. We shall add them in Section~\ref{se:5}.}
\eq{
I_{D} = -\frac{1}{\kappa_D}\,\int_{\MM} \nts \nts \extd^{D} x \sqrt{|g^{(D)}|} \,R^{(D)}\,,
}{eq:2de2}
and make a spherically symmetric Ansatz for the line-element,
\eq{
\extd s^2 = g_{\mu\nu}^{(D)}\extd x^\mu\extd x^\nu = g_{\alpha\beta}\extd x^\alpha \extd x^\beta + \frac{1}{\lambda}\,\phi^{2/(D-2)}\,\extd \Om^2_{S_{D-2}}\,,
}{eq:2de3}
where $\{\mu,\nu\}\!\!:$ $\{1,\dots,D\}$, $\{\al,\be\}\!\!:$ $\{1,2\}$, $\extd \Om^2_{S_{D-2}}$ is the line-element of the round $(D-2)$-sphere and $\lambda$ is a parameter with dimension of inverse length squared, which renders the scalar field $\phi$ dimensionless. The latter -- often called ``dilaton'' -- and the 2-dimensional metric $g_{\alpha\beta}$ both depend on the two coordinates $x^\alpha$ only. We parameterize the dimension by $D=2+\eps$, with the intention to take $\eps\to 0$ in the end, but for the time being $\eps$ need not be small. 

Inserting the Ansatz \eqref{eq:2de3} into \eqref{eq:2de2} and integrating over the angular part obtains a 2-dimensional dilaton gravity action 
\eq{
I_{\rm 2dg}^\eps = -\frac{1}{\kappa}\,\int_{\MM} \nts \nts \extd^2 x \sqrt{|g|} \, \left(\phi R-\frac{1-\eps}{\eps \phi}\,(\nabla\phi)^2-\lambda \eps (1-\eps)\phi^{1-2/\eps}\right)\,,
}{eq:2de4}
with a gravitational coupling constant given by
\eq{
\kappa=\frac{\kappa_{2+\eps} \lambda^{\eps/2}}{A_{S_{\eps}}} = \kappa_{2+\eps} \lambda^{\eps/2}\,\frac{\Gamma(\frac12+\frac{\eps}{2})}{2\pi^{\frac{1}{2}+\frac{\eps}{2}}}\,,
}{eq:2de5}
where the surface area $A_{S_\eps}$ comes from integration over the $\eps$-dimensional unit sphere. The appearance of the dimensionful constant $\lambda$ in the action and the coupling constant is a consequence of the Ansatz \eqref{eq:2de3}. It will survive the limit and arises because in $2+\eps$ dimensions Newton's constant is dimensionful.

So far we have not achieved much: the limit $\eps\to 0$ of \eqref{eq:2de4} still either is undefined or trivial, even after suitable rescalings of $\phi$, $\kappa$ and $\lambda$. This is so, because any rescaling that makes the kinetic term in \eqref{eq:2de4} finite automatically scales the $\phi R$-term to zero. Even a dilaton-dependent Weyl rescaling does not help: the conformal factor becomes singular in the limit, so calculating quantities for conformally related models and taking the limit $\eps\to 0$ there is possible, but leads to singular quantities in the original formulation and therefore is pointless. However, we note that the solutions for the metric to the equations of motion following from \eqref{eq:2de4},
\eq{
\extd s^2 = \Big(\lambda - 2a\, r^{1-\eps}\Big)\extd \tau^2 + \Big(\lambda - 2a \, r^{1-\eps}\Big)^{-1}\extd r^2\,,
}{eq:2de7}
possess a well-defined limit. The constant of motion $a$ essentially is the ADM mass. We have used Euclidean signature and diagonal gauge to represent \eqref{eq:2de7}, but of course our statements are gauge independent. Since the limit $\eps\to 0$ of \eqref{eq:2de7} is accessible there is a chance that another action -- leading to the same solutions \eqref{eq:2de7} -- permits a meaningful limit.

With this in mind we exploit now a duality discovered in \cite{Grumiller:2006xz}. The original action \eqref{eq:2de4} leads to the same 2-parameter family of line-elements \eqref{eq:2de7} as solutions of the classical equations of motions as the dual action
\eq{
\tilde{I}_{\rm 2dg}^\eps =-\frac{1}{\tilde{\kappa}}\,\int_{\MM} \nts \nts \extd^2 x \sqrt{|g|} \, \left(\tilde{\phi} R - 2a (1-\eps) \tilde{\phi}^{-\eps}\right)\,.
}{eq:2de6}
 In the original formulation \eqref{eq:2de4} $\lambda$ is a dimensionful parameter in the action and $a$ emerges as constant of motion. In the dual formulation \eqref{eq:2de6} the respective roles are reversed. The dual dilaton field is related to the original one by $\tilde{\phi}=\phi^{1/\eps}$. The dual coupling constant $\tilde{\ka}$ is arbitrary. 

It is straightforward to take the limit $\eps\to 0$ of \eqref{eq:2de6}.
\eq{
\tilde{I}_{\rm C} := \lim_{\eps\to 0}\tilde{I}^\eps_{\rm 2dg} = -\frac{1}{\tilde{\kappa}}\,\int_{\MM} \nts \nts \extd^2 x \sqrt{|g|} \, \big(\tilde{\phi} R - 2a \big)
}{eq:2de9}
The action $\tilde{I}_{\rm C}$ coincides with the geometric part of the CGHS action \cite{Callan:1992rs}. Thus, we have succeeded to obtain a well-defined non-trivial (dual) limit. Because the duality is involutive \cite{Grumiller:2006xz}, we now dualize $\tilde{I}_{\rm C}$ and obtain in this way the result we are seeking. The action dual to $\tilde{I}_{\rm C}$ is given by
\eq{
I_{\rm L} = -\frac{1}{\hat{\kappa}}\,\int_{\MM} \nts \nts \extd^2 x \sqrt{|g|} \, \big(\phi R - (\nabla\phi)^2 - \lambda e^{-2\phi} \big)\,.
}{eq:2de10} 
The coupling constant $\hat{\kappa}$ is arbitrary, and we shall exhibit its relation to $\kappa$ in Section \ref{se:4}. The action \eqref{eq:2de10} is recognized as the Liouville action \eqref{eq:2de34}.
Therefore, following the l'Hospital rule established in this Section, the limiting action \eqref{eq:2de1} is the Liouville action $I_L$. This is our main result.

We discuss now some features of \eqref{eq:2de10}. The range of the dilaton $\phi$ naturally is $(-\infty,+\infty)$. This, however, is not the case for spherically reduced gravity \eqref{eq:2de4} where $\phi$ must be non-negative. We shall see the geometric reason for this extension of the range in the next Section. An interesting property of \eqref{eq:2de10} is the invariance of the scalar field equation under local Weyl rescalings,
\eq{
g_{\mu\nu} \to e^{2\si}g_{\mu\nu}\,,\qquad \phi\to\phi+\si\,.
}{eq:2de32}
The action \eqref{eq:2de10} also is invariant (up to boundary terms and rescalings of $\lambda$) independently under global Weyl rescalings and constant shifts of the dilaton,
\eq{
g_{\mu\nu} \to e^{2\si_0}g_{\mu\nu}\,,\qquad \phi\to\phi+\si_1 \,.
}{eq:2de37}
With the redefinitions $\varphi = -2\phi$ and $m^2=2\lambda$ we can represent the action \eqref{eq:2de10} as
\eq{
I_L = \frac{1}{2\hat{\kappa}}\,\int_{\MM} \nts \nts \extd^2 x \sqrt{|g|} \, \big(\varphi R + \frac12 (\nabla\varphi)^2 + m^2 e^{\varphi} \big)\,.
}{eq:2de11}
Up to notational differences this coincides with the Liouville action considered e.g.~in \cite{Jackiw:2005su}, where further properties of \eqref{eq:2de10} are discussed. 
For $m=0$ our action \eqref{eq:2de11} coincides with the one proposed by Mann and Ross \cite{Mann:1992ar} as the $D\to 2$ limit of General Relativity (cf.~also \cite{Lemos:1993hz}). Their construction employs a Weyl transformation in $2+\eps$ dimensions ($\square:=g^{\mu\nu}\nabla_\mu\nabla_\nu$) 
\eq{
\tilde{g}_{\mu\nu}=e^\Psi g_{\mu\nu}\,,\qquad \tilde{R} = e^{-\Psi}\big(R-(1+\eps)\square\Psi-\frac{\eps(1+\eps)}{4} (\nabla\Psi)^2\big)\,,
}{eq:Mann1}
an ad-hoc subtraction between original and transformed action
\eq{
I_{\rm MR} := -\frac{1}{\ka_{2+\eps}} \int_{\MM} \nts \nts \extd^{2+\eps}x \,\big(\sqrt{|g|}R-\sqrt{|\tilde{g}|}\tilde{R}\big)\,,
}{eq:Mann2}
and the same rescaling of the gravitational coupling constant employed in the present work,
\begin{align}
I_{\rm MR} &= \frac{\eps}{2\ka_{2+\eps}} \int_{\MM} \nts \nts \extd^{2+\eps}x \sqrt{|g|}\,\big(\Psi R +\frac{1}{2} (\nabla\Psi)^2 + {\mathcal O}(\eps)\big)
\label{eq:Mann3} \\
& \to \frac{1}{2\hat{\ka}} \int_{\MM} \nts \nts \extd^2 x \sqrt{|g|}\,\big(\Psi R +\frac{1}{2} (\nabla\Psi)^2\big) \;{\rm as}\;\eps\to 0\,.
\label{eq:Mann4}
\end{align}
Thus their $\eps\to 0$ limiting action \eqref{eq:Mann4} differs from our \eqref{eq:2de10} in that the Liouville exponential is missing.

As mentioned in the introduction the metric $g$ and scalar field $\phi$ in \eqref{eq:2de10} are dynamical fields. For sake of completeness and to avoid confusion we recall here another interpretation of \eqref{eq:2de10}. In the approach of \cite{Distler:1989jt} the emergence of the Liouville action basically comes about as follows. Starting point is the path integral for bosonic strings with flat target-space metric,
\eq{
Z=\int{\mathcal D}g{\mathcal D}X \,e^{-\frac{1}{8\pi} \int \extd^2x\sqrt{|g|}(g^{\mu\nu}\partial_\mu X^a\partial_\nu X^b \eta_{ab} + \mu_0)}\,,
}{eq:2de44} 
where the measure contains the ghost- and gauge-fixing part. The kinetic term for the target-space coordinates is classically invariant under Weyl rescalings of the world-sheet metric $g\to e^{2\si} g$, but the measure is not,
\eq{
{\mathcal D} X \to {\mathcal D} X \,e^{I_L}\,,
}{eq:2de45}
where $I_L$ is the Liouville action \eqref{eq:2de10}, with $\phi$ replaced by $-\sigma$ (the coupling $\hat{\ka}$ depends on the dimension $d$ of the target-space). Analogous considerations apply to the ghost measure, so that for critical strings, $d=26$, the Liouville contribution to the action generated by \eqref{eq:2de45} cancels. For non-critical strings, $d\neq 26$, the Liouville contribution survives and is crucial to restore conformal invariance at the quantum level.\footnote{Actually, the full story is more complicated, involves a conjecture and eventually leads to \eqref{eq:2de33}, which generalizes \eqref{eq:2de10}; cf.~\cite{Nakayama:2004vk} for a review.} We emphasize that in the approach mentioned in this paragraph the metric (gauge-fixed to conformal gauge) is non-dynamical, while the field $\phi$ plays the role of the conformal factor.\footnote{Alternatively, it is also possible to interpret \eqref{eq:2de33}, supplemented by a kinetic term for the target-space coordinates, as a sigma model, i.e., a critical string theory. Then $\Phi$ is the dilaton, $g$ is the world-sheet metric, the target-space metric is flat and the term proportional to $\mu$ comes from a non-trivial tachyon background, cf.~e.g.~\cite{Nakayama:2004vk}. Another possibility was studied in \cite{Bergamin:2004pn}: the metric is dynamical and a specific Liouville action of type \eqref{eq:2de33} arises directly from the Polyakov action upon first integrating out the target-space coordinates, and then 'integrating in' an auxiliary scalar field $\Phi$.} This is quite different from the interpretation of \eqref{eq:2de10} in the present work, where metric and scalar field are dynamical.

\section{Consistency checks}\label{se:4}

\subsection{Geometric properties}

The classical solutions of the equations of motion descending from the Liouville action \eqref{eq:2de10} are given by
\eq{
\extd s^2 = \big(\lambda-2a\,r\big)\extd \tau^2 + \big(\lambda-2a\,r\big)^{-1}\extd r^2\,,\quad \phi = \ln{r}
}{eq:2de24}
This is compatible with \eqref{eq:2de7} in the $\eps\to 0$ limit. It should be noted that the coordinates in \eqref{eq:2de24} and \eqref{eq:2de7} have somewhat unusual physical dimensions: $\tau$ has dimension of length squared and $r$ is dimensionless. Exactly like \eqref{eq:2de7} the line-element \eqref{eq:2de24} exhibits a Killing vector $\partial_\tau$. The Killing norm is normalized to $\lambda$ at $r=0$, which may be considered as an ``asymptotic boundary'' because $\phi$ tends to $-\infty$. This brings us back to the issue of the range of $\phi$. It is now clear why the limiting geometry \eqref{eq:2de24} implies that $\phi$ can be also negative: if we restricted $\phi$ to positive values, we would impose a cutoff $r_0=1$ on the radial coordinate $r$. But this cut-off would be artificial, as neither geometry nor dilaton field exhibit any pathological behavior there. Only by allowing $\phi\in(-\infty,\infty)$ is it possible to achieve $r\in(0,\infty)$. There is a subtlety regarding singularities: even for arbitrarily small $\eps$ the line-elements \eqref{eq:2de7} have a curvature singularity at $r=0$. The limiting solution \eqref{eq:2de24}, however, does not exhibit any curvature singularity. Instead, it is the dilaton field that becomes singular as $r=0$ is approached. 

We assume from now on that $\lambda$ and $a$ are positive. 
Then for Lorentzian signature there is a Killing horizon at $\phi=\phi_h$, where
\eq{
\phi_h = \ln{\frac{\lambda}{2a}}\,.
}{eq:2de25}
This is consistent with \eqref{eq:2de7} which for Lorentzian signature also exhibits a Killing horizon for positive $\lambda$ and $a$. The constant of motion $a$ here plays the role of a Rindler acceleration, except that its dimension is one over length squared rather than one over length. The associated Unruh temperature can be derived in various standard ways, e.g.~from surface gravity. The result is
\eq{
T_{\rm U}= \frac{a}{2\pi}\,\lambda^{-1/2}\,.
}{eq:2de26}
The appearance of $\lambda^{-1/2}$ guarantees that the Unruh temperature $T_{\rm U}$ has the correct dimension of one over length. The same result can also be obtained from calculating the Hawking temperature associated with \eqref{eq:2de7} for positive $\eps$, and taking the limit $\eps\to 0$ in the end.

We have thus demonstrated that the limiting solutions \eqref{eq:2de24} are consistent with the family of solutions \eqref{eq:2de7}.

\subsection{Does matter curve geometry?}

According to standard folklore ``matter tells geometry how to curve''. In particular, in the absence of matter spacetime should be Ricci-flat. 
\eq{
R_{\mu\nu} = 0 \qquad \Leftrightarrow\qquad  T_{\mu\nu} = 0
}{eq:2de15}
This is certainly the case for Einstein gravity. Consistently, it should be true also for the limiting action \eqref{eq:2de10}. 

There is a complication, however. We have to consider the dilaton as part of the geometry, because the result \eqref{eq:2de10} arises as the limit \eqref{eq:2de1} of a purely geometric action. Therefore, by ``matter'' we always refer to some additional (physical) degrees of freedom, like scalar or Fermi fields, which we denote by $\psi_i$. The total bulk action is given by
\eq{
I_{\rm tot}[g,\phi,\psi_i] = I_{\rm L}[g,\phi] + I_{\rm mat}[g,\phi,\psi_i]\,,
}{eq:2de13}
and the energy momentum tensor is constructed from $I_{\rm mat}$ in the usual way,
\eq{
T_{\mu\nu} = \frac{2}{\sqrt{|g|}}\,\frac{\de I_{\rm mat}}{\de g^{\mu\nu}}\,.
}{eq:2de14} 
Its trace is denoted by $T:=T^\mu{}_\mu$. If the matter action depends on the dilaton we also need the definition
\eq{
\hat{T} = \frac{1}{\sqrt{|g|}}\,\frac{\de I_{\rm mat}}{\de \phi}\,.
}{eq:2de16}

We investigate now whether we can reproduce \eqref{eq:2de15} for generic dilaton gravity \eqref{eq:2de17}, supplemented by some matter action analog to \eqref{eq:2de13}-\eqref{eq:2de16}.
The equations of motion are
\begin{align}
R+\partial_\phi U (\nabla\phi)^2+2U\square\phi-\partial_\phi V &= \hat{\ka} \hat{T}\,,\\
2U \nabla_\mu\phi\nabla_\nu\phi - g_{\mu\nu} U(\nabla\phi)^2-g_{\mu\nu} V + 2\nabla_\mu\nabla_\nu\phi-2g_{\mu\nu}\square \phi &= -\hat{\ka} T_{\mu\nu}\,. 
\label{eq:2de42}
\end{align}
The trace of the second equation simplifies to
\eq{
2V+2\square\phi = \hat{\ka} T\,,
}{eq:2e18}
which allows to express the first equation as
\eq{
R = \hat{\ka} \hat{T} - \hat{\ka} U T + 2UV - \partial_\phi U(\nabla\phi)^2+\partial_\phi V\,.
}{eq:2e19}
In two dimensions a spacetime is Ricci-flat if and only if the Ricci scalar vanishes. Therefore, in the absence of matter ($\hat{T}=T=0$) spacetime is Ricci-flat if and only if
\eq{
2UV - \partial_\phi U(\nabla\phi)^2+\partial_\phi V = 0\,.
}{eq:2e20}
The condition \eqref{eq:2e20} holds only for a very specific class of dilaton gravity models. It is gratifying that the Liouville action \eqref{eq:2de10} [$U_{\rm L}=1$, $V_{\rm L}= \lambda e^{-2\phi}$] belongs to this class. Consistently, also its dual, the CGHS action \eqref{eq:2de9} [$U_{\rm C}(\phi)=0$, $V_{\rm C}(\phi)=\rm const.$], belongs to this class. 

If matter is not coupled to the dilaton then $\hat{T}=0$ and \eqref{eq:2e19} for Liouville gravity can be represented as
\eq{
R = -\hat{\ka} \, T\,.
}{eq:2de21}
This is as close an analog of Einstein's equations as possible in two dimensions \cite{Teitelboim:1984}. The simplest example of an energy-momentum tensor is $T_{\mu\nu}=-g_{\mu\nu} \Lambda/\hat{\ka}$, which just amounts to the addition of a cosmological constant to $I_L$:
\eq{
I_{\rm L\La} = -\frac{1}{\hat{\kappa}}\,\int_{\MM} \nts \nts \extd^2 x \sqrt{|g|} \, \big(\phi R - (\nabla\phi)^2 - \lambda e^{-2\phi} - \La\big)
}{eq:2de30}
The equations of motion yield
\eq{
R = 2\La\,.
}{eq:2de31}
Such spacetimes are maximally symmetric, i.e., they exhibit three Killing vectors.

We have thus demonstrated that the limiting action \eqref{eq:2de10} is consistent with the Einsteinian relation \eqref{eq:2de15}. Matter indeed ``tells geometry how to curve''.

\subsection{Entropy and coupling constant}

The Bekenstein-Hawking entropy is determined by the dilaton evaluated at the horizon \cite{Gegenberg:1994pv}. For Liouville gravity \eqref{eq:2de10} we obtain
\eq{
S = \frac{4\pi}{\hat{\ka}}\ln{\frac{\lambda}{2a}} = S_0 - \frac{4\pi}{\hat{\ka}} \ln{a}\,.
}{eq:2de23}
Here $S_0$ is independent from $a$, the only constant of motion.
On the other hand, the Bekenstein-Hawking entropy for the Schwarzschild black hole in $2+\eps$ dimensions is given by
\eq{
\tilde{S} = \frac{4\pi}{\ka_{2+\eps}}M^{\eps/(\eps-1)} = \tilde{S}_0 - \frac{2\pi\eps}{\ka}\ln{M} + {\mathcal O}(\eps^2)\,,
}{eq:2de27}
where $\tilde{S}_0$ is independent from $M$, the only constant of motion, and $\ka$ is defined in \eqref{eq:2de5}. Note that the physical units of $M$ are irrelevant here because any change of mass units only shifts $\tilde{S}_0$.

The expressions \eqref{eq:2de23} and \eqref{eq:2de27} are essentially equivalent upon relating the respective constants of motion by $a\propto M$. To achieve quantitative agreement we must identify
\eq{
\eps \hat{\ka} = 2\ka = \lim_{\eps\to 0} \ka_{2+\eps} \,.
}{eq:2de28}
For any finite choice of $\hat{\ka}$ the relation \eqref{eq:2de28} is perfectly consistent with our starting point, the assumption that $\kappa_{2+\eps}$ in \eqref{eq:2de1} scales with $\eps$. The factor 2 in the middle equation \eqref{eq:2de28} comes from the 0-sphere (which consists just of two points).

We have thus demonstrated that the limit \eqref{eq:2de1} with $\ka_{2+\eps}\propto \eps$ is consistent with the behavior of entropy \eqref{eq:2de23}.

\subsection{Miscellaneous further checks}

In our paper on the duality \cite{Grumiller:2006xz} we considered in detail a 2-parameter family of actions that included spherically reduced gravity from any dimension $D$. We found there that the dual model is conformally related to spherically reduced gravity from a dual dimension $\tilde{D}=(2D-3)/(D-2)$. For $D\to 2$ from above we obtain $\tilde{D}\to \infty$. Thus, for consistency our Liouville gravity action \eqref{eq:2de10} should be dual to a model that is conformally related to spherically reduced gravity from $\tilde{D}=\infty$ dimensions. But we know already that \eqref{eq:2de10} is dual to \eqref{eq:2de9}, so it remains to be shown that \eqref{eq:2de9} is conformally related to spherically reduced gravity from $\tilde{D}=\infty$ dimensions. Taking the limit $\eps\to\infty$ in \eqref{eq:2de4} and rescaling $\la$ appropriately yields an action of type \eqref{eq:2de17} with $U_{\rm W}=-1/\phi$ and $V_{\rm W}\propto \phi$. This is recognized as the ``Witten black hole'' action \cite{Witten:1991yr}, and it is indeed conformally related to the CGHS action \eqref{eq:2de9}.

%... note: for the $AdS$ case the action \eqref{eq:2de30} is conformally related to the dual action that emerges from the Kaluza-Klein reduction of conformally flat spaces from 4 to 3 dimensions ($U=-\tanh{(\phi/2)}/2$, $V\propto\sinh{\phi}$), cf.~Eq.~(48b) in \cite{Grumiller:2006ww}; if that relation is of interest I can explain the details here ...

%{\bf Now comes a half-baked paragraph on asymptotic safety:} ... in \cite{Kawai:1992fz}: perturbative quantization of gravity in $2+\eps$ dimensions; they separate conformal mode, find a UV fixed point (for $c<25$), calculate anomalous dimension of the cosmological constant term (their series for the anomalous dimension is of the form $1/2+1/4+\mathcal{O}(1/8)$ for $c=1$ and thus converges slowly; for $c>12$ it does not converge at all); they argued that the limit $\eps\to 0$ corresponds to the strong coupling phase of gravity in $2+\eps$ dimensions rather than to gravity in $2+\eps$ dimensions at its fixed point; the dominant counter term that they found was the Liouville action \eqref{eq:2de11} for $m=0$.  ... OTHER LITERATURE? cf.~\cite{Ginsparg:1993is} ...

%... further discussion? ...

\section{Boundary terms and summary}\label{se:5}

%We have succeded to construct a meaningful and non-trivial limit \eqref{eq:2de1}. 
We extend now our results as to include boundary terms, i.e., we consider the limit
\eq{
\Gamma_{\rm L}:=\lim_{\eps\to 0} \Gamma_{2+\eps}=\lim_{\eps\to 0}-\frac{1}{\kappa_{2+\eps}} \left[\int_{\MM} \nts\nts \extd^{2+\eps} x \sqrt{|g|} \,R +2\int_{\dM} \nts\nts \extd^{1+\eps} x \sqrt{|\ga|} \,K\right]\,.
}{eq:2de1b}
Again $\kappa_{2+\eps}$ is supposed to scale proportional to $\eps$; $\MM$ is a Riemannian manifold in $2+\eps$ dimensions with metric $g_{\mu\nu}$ and $R$ is the Ricci scalar; $\dM$ is the boundary of $\MM$ with induced metric $\gamma_{\mu\nu}$ and $K$ is the extrinsic curvature. Rather than attempting to perform similar steps as above for the boundary action we take a shortcut. Since we know already that the correct bulk action is given by \eqref{eq:2de10} we simply supplement the latter by the appropriate boundary terms. They comprise the dilaton gravity analog of the Gibbons-Hawking-York boundary term and a Hamilton-Jacobi counterterm \cite{Grumiller:2007ju}. The full limiting action is given by 
\begin{multline}
\Gamma_{\rm L} = -\frac{1}{\hat{\kappa}}\,\int_{\MM} \nts \nts \extd^2 x \sqrt{|g|} \, \big(\phi R - (\nabla\phi)^2 - \lambda e^{-2\phi} \big) \\
-\frac{2}{\hat{\kappa}}\, \int_{\dM} \bns dx \, \sqrt{\gamma}\,X\,K + \frac{2}{\hat{\kappa}}\, \int_{\dM} \bns dx \,\sqrt{\gamma}  \, \sqrt{\lambda} \, e^{-\phi} \,.
\label{eq:2de12}
\end{multline}

Our procedure is summarized in the following diagram:\footnote{The only step not discussed so far is the one called ``oxidation'', the inverse procedure of ``reduction''. Since reduction means integrating out the $\eps$-sphere, oxidation after taking the limit $\eps\to 0$ implies ``integrating in'' the 0-sphere. But this just amounts to rescaling the coupling constant $\hat{\ka}$ by a factor of 2 and therefore is trivial.}
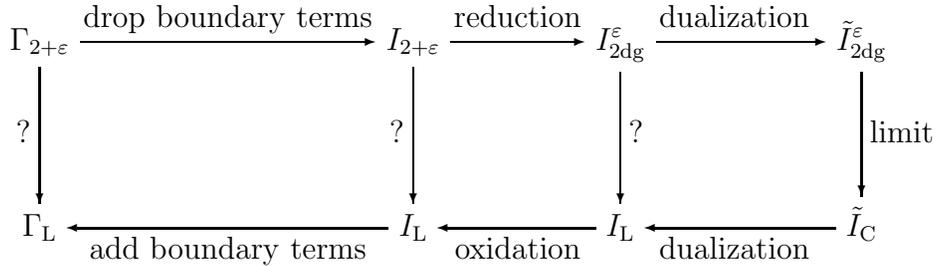
\begin{diagram}%[leftflush]
\Ga_{2+\eps} & \rTo^{\rm drop\,\,boundary\,\,terms} & I_{2+\eps} & \rTo^{\rm reduction} & I_{\rm 2dg}^{\eps} &  \rTo^{\rm dualization} &  \tilde{I}_{\rm 2dg}^{\eps} \\
\dTo^{?} & & \dTo^{?} & & \dTo_{?}  & & \dTo_{\rm limit} \\
\Gamma_{\rm L} & \lTo_{\rm add\,\,boundary\,\,terms} & I_{\rm L} & \lTo_{\rm oxidation} & I_{\rm L} & \lTo_{\rm dualization} & \tilde{I}_{\rm C} \\
\end{diagram}
%\label{p:1}
The arrows decorated with a question mark refer to our inability to construct directly a meaningful limit $\eps\to 0$. Therefore, we started with the upper left corner, proceeded with the steps indicated above the arrows to the upper right corner, then took the limit to the lower right corner, and finally inverted all steps (as indicated below the arrows) to arrive at the lower left corner.

Now it is evident how the various approaches to gravity near two dimensions [Eq.~\eqref{eq:2de1}] or in two dimensions [Eqs.~\eqref{eq:2de17}-\eqref{eq:2de34}] are connected. Thus we conclude that the closest analog to the Einstein-Hilbert action in two dimensions is Liouville gravity \eqref{eq:2de12}.

\section*{Acknowledgments}

DG thanks Robert Mann for discussions.

This work is supported in part by funds provided by the U.S. Department of Energy (DOE) under the cooperative research agreement DEFG02-05ER41360.
DG has been supported by the Marie Curie Fellowship MC-OIF 021421 of the European Commission under the Sixth EU Framework Programme for Research and Technological Development (FP6). 

%\bibliographystyle{fullsort}
%\bibliography{review}

\providecommand{\href}[2]{#2}\begingroup\raggedright\endgroup

\end{document}